\documentstyle[aps,prb,twocolumn,psfig]{revtex}
\begin{document}
\title{Optical nonlinearity due to intersubband transitions in semiconductor 
quantum wells}

\author{Sanghamitra Mukhopadhyay\cite{oldaff}}
\address{Inter University Consortium for DAE Facilities, Khandwa Road, Indore
452 017, India}

\author{K. C. Rustagi}
\address{Laser Physics Division, Centre for Advanced Technology,
Indore 452 013, India.}
\maketitle

\begin{abstract}
\it{
We have calculated the contribution of intersubband transitions to
the third order optical nonlinear susceptibility,
$\chi^{(3)}(\omega,\omega,\omega)$ for nonresonant as
well as resonant third harmonic generation and 
$\chi^{(3)}(\omega,-\omega,\omega)$ for nonlinear refraction and absorption.
As examples, we consider InAs/AlSb and GaAs/GaAlAs quantum wells. The effects
of finite barrier height, energy band nonparabolicity, and high 
carrier concentrations are included. It is shown that quantum confinement, 
rather than the
band nonparabolicity, is responsible for high values of nonresonant 
$\chi^{(3)}_{zzzz}$. Very high values of $\chi^{(3)}_{zzzz}$ are obtained 
for third harmonic generation and two photon absorption for incident
wavelength near 10.6 $\mu$m. Intensity dependence of refractive index
and of absorption co-efficient is also
discussed for intensity well above the saturation intensity. Effective medium
theory is used to incorporate the collective 
effects.
}
\end{abstract}


\section{Introduction}
Intersubband transitions (ISBT) in semiconductor quantum wells (QW) have
attracted a lot of attention \cite{wes,fej,hey} in 
nonlinear optics since the work of West and Eglash \cite{wes}. As
emphasized in a recent review by Almogy and Yariv \cite{alm}, the main reason
for this interest is the occurrence of narrow transitions with large 
oscillator strengths. 
In bulk semiconductors mobile carrier optical nonlinearity, $\chi^{(3)}$ due 
to intraband transitions is proportional to $\frac{1}
{\omega^4}\frac{\partial^4 E}{\partial k^4}$ at low frequencies 
\cite{but1,jha}, where photon energy $\hbar \omega \ll E_g$, the 
semiconductor bandgap; $E$ is the electron energy,
and $k$ is the electron wave vector. This suggests that nonlinearity in bulk
semiconductors increases with increase in the energy band nonparabolicity. 
Some attempts \cite{blo,cha,cho} have been made to extend the above relation 
to calculate $\chi^{(3)}$ due to mobile carriers in a semiconductor QW with a
view to enhance the nonparabolicity and hence the optical nonlinearity.
However, as noted by Jha and Bloembergen \cite{jha}, 
the low frequency limit is valid when 
$\hbar \omega $ is much smaller than the relevant bandgaps, which in the 
present case is the intersubband gap, $E_{ISBT}$. Thus in a superlattice (SL)
or a multiple quantum well (MQW) there
are two relevant low-frequency regimes: $E_{ISBT} \le \hbar \omega < E_g$,
and second, $\hbar \omega < E_{ISBT}$. For the former regime, which is of more
practical interest, the above
simple relation is not applicable and it is important to understand the role
played by quantum confinement \cite{mar1,mar2}. Resonant enhancement of
nonlinearity at ISBT is the second major theme in the exploitation of 
quantum wells as nonlinear materials. High values of $\chi^{(3)}$ have been
reported by third harmonic generation (THG) involving ISBT resonances 
\cite{sir,lie}. The third important aspect of optical nonlinearities
involving ISBT is the significant effects of electron-electron interaction.
Two of the most important consequences of including the Coulomb interaction 
are the shift of the intersubband resonance due to depolarization effects
\cite{and,khu2,war,zal} and the predicted occurence of optical bistability 
\cite{prlpap,new}. Most of the work cited above has emphasized one or the 
other
aspect. In this paper we develope a theory of the optical nonlinearities
in the intersubband transition region including the essential complexities
such as, finite barrier height which allows the electron wavefunctions
to extend into the barrier region, nonparabolicity which affects the
intersubband energies, the subband effective masses, and the collective effect
which is important at high carrier concentrations. To do this we first
calculate the single particle response including realistic subband dispersions
and then include electron-electron interaction through an effective medium 
theory.

In Section II we present the theoretical formulation.
Various approximations necessary to deal with nonparabolic bands in quantum 
wells are stated as well the necessary details of the expressions used for
evaluating nonlinear susceptibilities. The nonlinear susceptibility 
$\chi^{(3)}_{zzzz}(-3\omega, \omega, \omega, \omega)$, here after denoted as
$\chi^{(3)}(3\omega)$ describing the THG and $\chi^{(3)}_{zzzz}(-\omega,
\omega, -\omega, \omega)$, hereafter denoted as $\chi^{(3)}(-\omega)$
are treated separately. To deal with resonant effects at high intensities
power broadening effects in resonances are also included. We also describe 
here the effective medium theory to account for depolarization terms due to 
electron-electron interaction.

In Section III the results of numerical calculations of THG involving ISBT
are discussed for InAs/AlSb MQW systems. This 
material system has become important because of its high barrier height 
(1.35 eV) which allows large charge density (of the order of 
10$^{19}$ cm$^{-3}$) to accumulate. Very high room temperature mobility 
(2 $\times$ 10$^4$ cm$^2$V$^{-1}$s$^{-1}$), and 
ultra-fast energy relaxation time (1 ps at room temperature) of the system
are also useful for device applications \cite{mar3}. We show a very high value
of $|\chi^{(3)}(3\omega)|$ (of the order of 10$^{-4}$ e.s.u. at 10.6 $\mu m$)
can be achieved from this system.

In Section IV NRA, $\chi^{(3)}(-\omega)$, are calculated for the same
material system and also for GaAs/GaAlAs MQW system. We find that large 
value of
$\chi^{(3)}(-\omega)$ can be achieved at two photon absorption (TPA)
wavelength which is near $\sim$ 10.6$\mu m$ for InAs/AlSb system. Since the 
values 
of Re$(\chi^{(3)}(-\omega))$ is larger than that of Im$(\chi^{(3)}(-\omega))$
near TPA wavelength useful applications of these systems in optical switching
devices may be expected. Optical bistability has already been predicted 
\cite{prlpap} for InAs/AlSb system after considering electron-electron 
interaction through the effective medium theory. This intrinsic optical 
bistability of a medium is of considerable theoretical interest.

In Section V the numerical results of intensity-dependent absorption
including the power-broadening term are presented. The calculated values 
obtained by using the effective medium theory to include the depolarization 
effects compare very
well with earlier experimental \cite{cra} and theoretical \cite{zal} results. 

The conclusions are presented in Section VI.

\section{Theoretical formulation of $\chi^{(3)}$}
In this section we first formulate the single particle response $\chi^{(3)}$
for ISBT, and then the electron-electron interaction is included through the
effective medium theory.

\subsection*{Single particle response}
We consider a MQW system consisting of rectangular wells and barriers of 
widths $L_W$ and $L_B$, respectively. The well region is
doped n-type so that at least the lowest subband is always populated. The $z$-
direction is chosen to be perpendicular to the plane of the wells with $z=0$ 
at the center of a well. The equation for the envelope function, $F(z)$ is 
then written as \cite{nag},
\begin{equation}
\frac{\hbar^2}{2m^{*}_{E_i}(E)} \frac{\partial^2F_i(z)}{\partial z^2} +
\left[ (E-E_{ci})-\frac{\hbar^2 k_{\perp}^2}{2m^{*}_{E_i}(E)} \right] F_i(z)
= 0 ,
\label{envf}
\end{equation}
and is solved with the continuity conditions,
\begin{mathletters}
\label{bcs}
\begin{equation}
F_W(z) = F_B(z),
\end{equation}
and
\begin{equation}
\frac{1}{m^{*}_{VW}(E)} \frac{dF_W(z)}{dz} = 
\frac{1}{m^{*}_{VB}(E)} \frac{dF_B(z)}{dz} 
\end{equation}
\end{mathletters}
for $|z|=L_W/2$.
In Eqs. \ (\ref{envf}) and (\ref{bcs}) $E_{ci}$ represents the band-edge 
energy, $E$ is the energy eigenvalue
including the component due to in-plane motion, $i$ stands for either 
the well (W) or the barrier (B), $\hbar k_{\perp}$ is the in-plane 
momentum of the electron; 
$m^{*}_{Ei}$ and $m^{*}_{Vi}$ are the energy and the velocity effective 
masses, respectively, and are given by 
\begin{mathletters}
\label{efms}
\begin{equation}
m^{*}_{E_i}(E) = m^{*}_{i}(0) \left\{ 1+ \alpha_i \left(E-E_{ci}\right) 
\right\}, 
\label{mei}
\end{equation}
and 
\begin{equation}
m^{*}_{V_i}(E) = m^{*}_{i}(0) \left\{ 1+ 2\alpha_i \left(E-E_{ci}\right) 
\right\}, 
\label{mvi}
\end{equation}
\end{mathletters}
where $m^{*}_{i}(0)$ is the band-edge effective mass in the bulk material,
and $\alpha_i$ is the nonparabolicity parameter of the $i$ layer.

We calculate the third order optical susceptibility, $\chi^{(3)}_{zzzz}$ in 
the $z$-direction. In this paper susceptibilities in 
the in-plane directions will not be considered. The expression for 
$\chi^{(3)}_{ijkl}(-\omega_{\sigma},\omega_1,\omega_2,\omega_3)$, where 
$\omega_{\sigma}=\omega_1+\omega_2+\omega_3$ can be obtained by using the 
standard density matrix perturbation formalism \cite{boy}, and has 48 terms 
which are reduced to 24 distinct terms for $i=j=k=l=z$.  

Resonant third order susceptibilities have been calculated for the cases of
THG and NRA by the sum-over-states method. For THG we get,
\begin{eqnarray}
\chi^{(3)}_{zzzz}(-3\omega,&\omega&,\omega,\omega) = \frac{e^4}
{\pi \hbar^5 L_W} \nonumber \\
&\times& \int_{0}^{\infty} \sum_{a} m^{*}_{V_W}(E) f(E) S(\omega) dE ,
\label{chi3thg}
\end{eqnarray}
where
$f(E)$ is the Fermi function given by
\begin{equation}
f(E) = \frac{1}{1+exp\{(E-E_a-E_f)/k_BT\}} ,
\label{fe}
\end{equation}
where $E$ is the energy corresponding to the in-plane momentum of the 
electron, $E_a$ is the energy 
eigenvalue of the state $a$ of the QW, and $E_f$ is the Fermi energy. The 
factor $S(\omega)$ in Eq. \ (\ref{chi3thg}) is 
\begin{eqnarray}
S(\omega) = &\;& \sum_{b \ne a} \sum_{c \ne a} \sum_{d \ne a} 
z_{ab} z_{bc} z_{cd} z_{da} \nonumber \\
&\times& \biggl\{ \frac{1}
{(\omega_{ba}-3\omega)(\omega_{ca}-2\omega)(\omega_{da}-\omega)} \nonumber \\
&+& \frac{1}{(\omega^{*}_{ba}+\omega)(\omega_{ca}-2\omega)(\omega_{da}-
\omega)} \nonumber \\
&+& \frac{1}{(\omega^{*}_{ba}+\omega)(\omega^{*}_{ca}+2\omega)(\omega_{da}
-\omega)} \nonumber \\
&+& \frac{1}{(\omega^{*}_{ba}+\omega)(\omega^{*}_{ca}+2\omega)
(\omega^{*}_{da}+3\omega)} \biggr\} \nonumber \\
&-& \sum_{b \ne a} \sum_{d \ne a} |z_{ab}|^2 |z_{ad}|^2  \nonumber \\
&\times& \frac{1}{2\omega} \biggl\{ \frac{1}
{(\omega_{da}-\omega)(\omega_{ba}-3\omega)} \nonumber \\
&+& \frac{1}{(\omega^{*}_{da}+3\omega)(\omega^{*}_{ba}+\omega)} \biggr\}
\label{sw}
\end{eqnarray}
where $z_{jk}$ are dipole matrix elements between two envelope states in the
$j$-th and $k$-th subbands. In Eq. \ (\ref{sw}), $\hbar\omega_{jk}=
E_{jk}-i\Gamma_{jk}$, where $E_{jk}$ is the intersubband energy
between the states $j$ and $k$, and $\Gamma_{jk}$ is the broadening parameter;
and $\omega^{*}_{jk}$ is the complex conjugate of $\omega_{jk}$. In general,
$z_{jk}$ and $E_{jk}$ depend on the in-plane momentum of the electron for
nonparabolic bands.
 
The NRA are the other important processes which
has not been discussed in detail in the literature \cite{khu}. The importance
of a complete theoretical expression of $\chi^{(3)}_{zzzz}(-\omega,\omega,
-\omega,\omega)$
was pointed out earlier by Bloembergen et. al. \cite{bloe} for a system where
close one and two photon resonances were present. In this article we present
a complete expression of $\chi^{(3)}_{zzzz}(-\omega,\omega,-\omega,\omega)$
due to ISBT for a heavily doped MQW system in which resonances are similar to
the case considered by Bloembergen et. al. \cite{bloe}
The third order susceptibility for NRA is given by
\begin{eqnarray}
\chi^{(3)}_{zzzz}(-\omega,\omega,-\omega,\omega)= &\;& \frac{e^4}
{3 \pi \hbar^5 L_W} \nonumber \\
&\times& \int_{0}^{\infty} \sum_a m^{*}_{VW}(E) f(E) T(\omega) dE ,
\label{chi3tpa}
\end{eqnarray}
where
\begin{eqnarray}
T(\omega) = &\;& \sum_{b \ne a} \sum_{c \ne a} \sum_{d \ne a} 
z_{ab} z_{bc} z_{cd} z_{da} \nonumber \\
&\times& \biggl\{ \frac{1}
{(\omega_{da}-\omega)\omega_{ca}(\omega_{ba}-\omega)} \nonumber \\
&+&\frac{1}{(\omega_{da}-\omega)\omega_{ca}(\omega_{ba}+\omega)} \nonumber \\
&+& \frac{1}{(\omega_{da}-\omega)(\omega_{ca}-2\omega)(\omega_{ba}-\omega)}
\nonumber \\
&+&\frac{1}{(\omega^{*}_{da}-\omega)(\omega_{ca}-2\omega)(\omega_{ba}-\omega)}
\nonumber \\
&+& \frac{1}{(\omega^{*}_{da}-\omega)\omega^{*}_{ca}(\omega_{ba}+\omega)}
\nonumber \\
&+& \frac{1}{(\omega^{*}_{da}+\omega)\omega^{*}_{ca}(\omega^{*}_{ba}+\omega)}
\nonumber \\
&+& \frac{1}{(\omega^{*}_{da}+\omega)\omega_{ca}(\omega_{ba}+\omega)}
\nonumber \\
&+& \frac{1}{(\omega^{*}_{da}+\omega)\omega_{ca}(\omega^{*}_{ba}-\omega)}
\nonumber \\
&+& \frac{1}{(\omega^{*}_{da}-\omega)\omega^{*}_{ca}(\omega_{ba}-\omega)}
\nonumber \\
&+& \frac{1}{(\omega^{*}_{da}+\omega)\omega^{*}_{ca}(\omega^{*}_{ba}-\omega)}
\nonumber \\
&+& \frac{1}{(\omega^{*}_{da}+\omega)(\omega^{*}_{ca}+2\omega)(\omega_{ba}+
\omega)} \nonumber \\
&+&\frac{1}{(\omega^{*}_{da}+\omega)(\omega^{*}_{ca}+2\omega)
(\omega^{*}_{ba}+\omega)} \biggr\} \nonumber \\
&-& \sum_{b} \sum_{d} |z_{ab}|^2 |z_{ad}|^2 \nonumber \\
&\times& \biggl\{\frac{1}
{(\omega_{da}-\omega)(\omega_{ba}-\omega)(\omega_{ba}-\omega)} \nonumber \\
&+& \frac{1}{(\omega_{da}-\omega)(\omega_{ba}+\omega)(\omega_{ba}+\omega)}
\nonumber \\
&+& \frac{1}{(\omega_{da}-\omega)(\omega_{ba}+\omega)(\omega_{ba}-\omega)}
\nonumber \\
&+& \frac{1}{(\omega^{*}_{da}-\omega)(\omega^{*}_{da}-\omega)(\omega^{*}_{ba}+
\omega)} \nonumber \\
&+& \frac{1}{(\omega^{*}_{da}+\omega)(\omega^{*}_{da}+\omega)(\omega^{*}_{ba}
+\omega)} \nonumber \\
&+& \frac{1}{(\omega^{*}_{da}+\omega)(\omega^{*}_{ba}+\omega)(\omega^{*}_{ba}+
\omega)} \nonumber \\
&+& \frac{1}{(\omega^{*}_{da}+\omega)(\omega_{ba}+\omega)(\omega_{ba}+\omega)}
\nonumber \\
&+&\frac{1}{(\omega^{*}_{da}-\omega)(\omega_{ba}-\omega)(\omega_{ba}-\omega)} 
\nonumber \\
&+& \frac{1}{(\omega^{*}_{da}-\omega)(\omega_{ba}-\omega)(\omega_{ba}+\omega)}
\nonumber \\
&+& \frac{1}{(\omega^{*}_{da}-\omega)(\omega^{*}_{da}-\omega)(\omega_{ba}-
\omega)} \nonumber \\
&+& \frac{1}{(\omega^{*}_{da}+\omega)(\omega^{*}_{da}+\omega)(\omega_{ba}
-\omega)} \nonumber \\
&+& \frac{1}{(\omega^{*}_{da}-\omega)(\omega^{*}_{da}+\omega)(\omega_{ba}+
\omega)} \biggr\}
\label{tw}
\end{eqnarray}
The terms following the double summations in the expressions for 
$S(\omega)$ and $T(\omega)$ are the seperated out terms containing the ground 
state as an intermediate state ($c=a$). The apparently divergent terms ($c=a$)
in the expression for $T(\omega)$ are reduced in the usual manner \cite{boy}.

It should be noted that the expression for $\chi^{(3)}(-\omega)$ is complete 
and is rigorously
calculated from the original 24 terms which are obtained from the 
perturbation theory in dissipative systems. These 24 terms then reduce to 12
terms due to $\omega_1=-\omega_2=\omega_3=\omega$. Since the imaginary part of
$\chi^{(3)}(-\omega)$ gives the two photon absorption (TPA) co-efficients, 
it should be positive for the entire frequency range. Taking the proper sign 
of the damping factor is, therefore, important. 
In the calculation of THG, the terms are 
calculated with the help of the perturbation theory in the absence of 
damping, and then the damping factors are introduced at the final expression
to obtain the Lorentzian broadening at $\omega$, 2$\omega$, and 3$\omega$
resonances \cite{but}.  It was shown \cite{hos} that in the THG dispersion,
the term containing $[(\omega_{ba}-3\omega)(\omega_{ca}-2\omega)
(\omega_{da}-\omega)]^{-1}$ is dominant among all the 48 terms and no 
cancellation of resonant terms is involved. Hence our approximation for 
$\Gamma_{ij}$ is justified. 

The nonresonant optical nonlinearity can be calculated from Eq. \
(\ref{chi3thg}) by putting $\Gamma=0$ taking the limit, $\omega\rightarrow 0$.
The resultant expression is 
\begin{eqnarray}
&\chi&^{(3)}_{zzzz}(0) = \frac{4e^4}{\hbar^5 \pi L_W}
\int_{0}^{\infty} \sum_{a} m^{*}_{V_W}(E) f(E) \nonumber \\
&\times& \left[ \sum_{b \ne a} \sum_{c \ne a} \sum_{d \ne a} 
\frac{z_{ab} z_{bc} z_{cd} z_{da}}{\omega_{ba} \omega_{ca} \omega_{da}}
- \sum_{b \ne a} \sum_{d \ne a} \frac{|z_{ab}|^2 |z_{da}|^2}{\omega_{da}
\omega_{ba}^2} \right] dE \nonumber \\
\;
\label{chi30}
\end{eqnarray}
In the infinite barrier height limit, the above expression may be 
cross-checked with the formula derived by
Rustagi and Ducuing \cite{rus1} for the particle-in-a-box model, which 
when suitably modified for an infinitely deep semiconductor QW takes the 
following form:
\begin{equation}
\chi^{(3)}_{zzzz}(0) = \frac{e^4 N_sm^{*3}_W L^9_W}{32 \hbar^6}
\sum_{n=1}^N \left(\frac{-2}{9 \pi^6 n^6} + \frac{140}{3 \pi^8 n^8}
-\frac{440}{\pi^{10} n^{10}} \right)
\label{chi0}
\end{equation}
where $N_s$ is the sheet carrier concentration, $m^*_W$ is the bulk bandedge 
effective mass and $N$ is the total number of occupied subbands.

To calculate the intensity-dependent absorption we have included power 
broadening following the general formalism \cite{boy} to obtain
the expression for the total (linear and nonlinear combined) 
optical susceptibility, $\chi (-\omega)$ as:
\begin{eqnarray}
&\chi& (-\omega) = -\frac{e^2}{\epsilon_0 L_W \pi \hbar^2} \nonumber \\
&\;& \int_{0}^{\infty} \sum_{a} \sum_{b} \left( m^{*}_{aV_W}(E) f_a(E)
-m^{*}_{bV_W}(E) f_b(E) \right) \nonumber \\
&\times& \frac{z_{ab} z_{ba} \left( \hbar\omega-E_{ba} 
-i\Gamma_2 \right) \Gamma^{-2}_2}{1 + \left(\hbar\omega-E_{ba} \right)^2 
\Gamma_2^{-2}+ 4 z_{ab}z_{ba}|E_W|^2 \left( \Gamma_1\Gamma_2 \right)^{-1}} dE
\label {2level}
\end{eqnarray}
where $a$ and $b$ are two consecutive energy levels $(b>a)$, $\Gamma_1=
\hbar/T_1$, and $\Gamma_2=\hbar/T_2$ with $T_1$, $T_2$ being the lifetime
and the dephasing time, respectively.

\subsection*{Electron-electron interaction}
It is generally accepted that the Coulomb interaction between the carriers
plays an important role in the enhancement of optical nonlinearity and the
blue shift of intersubband resonances. For many systems, e.g. a metallic
sphere, or a metal-insulator interface simple electromagnetic theories of
surface plasmons and surface polaritons have been important in understanding 
this interaction \cite{bre}. The effective medium theory is a similar one 
which we will
apply for MQWs in this section. The direct physical nature of this theory
allows us to predict the plasmon frequency-dependence on physical parameters,
such as the width and the dielectric constant of the barrier material.
For low carrier concentration the modification from single electron
approximation is not so important, on the
other hand, for high carrier concentrations collective effects are
important \cite{and,khu2,new}. We have already shown that the effective 
medium theory can give a good representation of the 
collective effects \cite{war} in linear absorption spectra of ISBT 
\cite{prlpap}. Here we extend the effective medium theory to nonlinear 
processes and 
show good agreements between our calculated results and experimental 
observations \cite{cra} as well as theoretical results \cite{zal} reported
earlier. These experimental results are otherwise explainable only when
many-electron calculations are done \cite{war}. This is an indication that
the effective medium theory is capable of representing the most important 
many-electron effects 
in such systems in a direct and physically attractive way. Recent 
calculations by Das Sarma and Hwang \cite{das} 
also show the importance of the barrier width in understanding the behavior
of plasmons in MQWs, which is in agreement with our results.

We take a MQW system as a layered medium with alternate layers of width $L_W=
\alpha d$ and $L_B=(1-\alpha) d$ where $d$ is the period of the MQW
and $\alpha$ is a fraction. For $d \ll \lambda$, the wavelength of light, the
composite medium can be described by an effective dielectric function,
$\tilde{\epsilon}_{ij}(\omega)$ \cite{bre}. The calculation of the nonlinear 
response has
to be performed in a self-consistent manner since the electric field in the 
well region, $E_W$ and in the barrier region, $E_B$ depend on the
dielectric functions as
\begin{mathletters}
\label{eweb}
\begin{equation}
E_W = \epsilon_B \tilde{E}/\left\{\epsilon_B \alpha + \epsilon_W
(1-\alpha) \right\},
\end{equation}
and
\begin{equation}
E_B = \epsilon_W \tilde{E}/\left\{\epsilon_B \alpha + \epsilon_W
(1-\alpha) \right\},
\end{equation}
\end{mathletters}
where $\tilde{E}$ corresponds to the average electric field in the 
$z$-direction given by,
\begin{equation}
\tilde{E} = \alpha E_W + (1-\alpha) E_B .
\label{etilde}
\end{equation}
The function $\epsilon_W$, the dielectric permittivity of the well layer, in 
turn, depends on $E_W$ according to
$ \epsilon_W = \epsilon_{bulk} + \epsilon_{intersubband}$ where $\epsilon_{
intersubband}$ is calculated from Eq. \ (\ref{2level}). For simplicity we 
have assumed that $\epsilon_B$, the dielectric permittivity of the barrier
layer is a constant in the frequency range of 
interest. In case of $z$-polarized light, the electric displacement vector 
across the interface is continuous, i.e.
\begin{equation}
\epsilon_W E_W = \epsilon_B E_B = \tilde{\epsilon} \tilde{E}.
\label{epswew}
\end{equation}
From Eqs. \ (\ref{eweb}), (\ref{etilde}), and (\ref{epswew}) we obtain 
the average dielectric constant of the composite medium as
\begin{equation}
\tilde{\epsilon} = \frac{\epsilon_W \epsilon_B}{\alpha \epsilon_B + (1-\alpha)
\epsilon_W} .
\label{epstilde}
\end{equation}

From Eqs. \ (\ref{2level}), (\ref{epswew}), and (\ref{epstilde}) we obtain the
following relation:
\begin{eqnarray}
P^2 C^2 x^3 &+& \Big[ 2 Re \left( PC \beta_0 e^{i\phi} \right) + 2P^2 CB
-yC^2 \Big] x^2   \nonumber \\
&+& \Big[ B^2 P^2 + \beta_0^2 + 2 Re \left( BP \beta_0 e^{i\phi} \right)
\nonumber \\
&-& 2BCy \Big] x - yB^2 = 0 ,
\label{cubic}
\end{eqnarray}
where
\begin{eqnarray*}
P &=& \frac{\alpha}{1-\alpha} \epsilon_B + \epsilon_{bulk} + 1 , \\
B &=& 1 + \left(\hbar\omega-E_{ba} \right)^2 \Gamma_2^{-2} , \\
C &=& 4 z_{ab}z_{ba} \left( \Gamma_1 \Gamma_2 \right)^{-1} , \\
\beta_0 e^{i\phi} &=& -\frac{e^2}{\epsilon_0 L_W}
\left( n_{sa}-n_{sb} \right) |z_{ba}|^2 \left( \hbar\omega-E_{ba} 
-i\Gamma_2 \right) \Gamma^{-2}_2 , \\
y &=& \Big| \frac{\epsilon_B}{1-\alpha} \tilde{E} \Big|^2 , \\
x &=& |E_W|^2 .
\end{eqnarray*}
In the above $n_{si}$ is the sheet carrier concentration of the $i$-th energy 
level.
Equation \ (\ref{cubic}) admits multivalued solutions for intensity in the 
well region as a function of the average intensity, $|\tilde{E}|^2$ for 
several values of physical parameters.

The value of the absorption co-efficient can be calculated from the imaginary
part of $\tilde{\epsilon}$ where the solution of Eq. \ (\ref{cubic}) is used
to calculate $\epsilon_W$. The value of the nonlinear refractive index for
large detuning or below the satuaration intensity can be calculated from 
\begin{equation}
\chi(-\omega) = \chi^{(1)}_{zz}(-\omega) + 3\chi^{(3)}_{zzzz}(-\omega)
|E_W|^2 ,
\end{equation}
where $\chi^{(1)}_{zz}(-\omega)$ is the linear susceptibility due to ISBT. In
this case, the collective effect can be introduced following the formulation 
as reported elsewhere \cite{prlpap}.

\section {Nonresonant $\chi^{(3)}_{\lowercase{zzzz}}(0)$}

The nonresonant third order nonlinearity $\chi^{(3)}_{zzzz}(0)$ due to ISBT 
has been calculated using Eq. \ (\ref{chi30}) for an InAs/AlSb MQW system 
used by Warburton et al. 
\cite{war}. The well and the barrier widths are 18 and 10 nm, respectively, 
and the carrier concentration is 2.61 $\times$ 10$^{12}$ cm$^{-2}$ at 10K 
whereby the second subband is populated. The values of the physical 
parameters for the calculations, and the results are given in Tables I and II,
respectively.
\begin{table}
\caption{Physical parameters used in the calculations}
\begin{tabular}{cccccc}
Material Systems&$m^{*}_W/m_0$&$m^{*}_B/m_0$&$\alpha_W$&$\alpha_B$&
$E_{CB}-E_{CW}$ \\
                &              &              &(eV$^{-1}$)&(eV$^{-1}$)&
(eV) \\ \hline
InAs/AlSb       &0.025&0.260&2.050&0.223&1.350 \\
GaAs/GaAlAs     &0.067&0.0919&0.573&0.295&0.261
\end{tabular}
$m_0$ is the free electron mass.
\end{table}

\begin{table}
\caption{ Calculated values of $\chi^{(3)}_{zzzz}(0)$ in e.s.u.}
\begin{tabular}{ccccc}
&Using Eq. \ (\ref{chi30}) & &\vline& Using Eq. \ (\ref{chi0}) \\
Parabolic& &Nonparabolic&\vline&   \\ \hline
-9.80 $\times$ 10$^{-7}$& &-5.08 $\times$ 10$^{-7}$&\vline&-5.27 $\times$ 
10$^{-9}$ 
\end{tabular}
\end{table}

The values of $\chi^{(3)}_{zzzz}(0)$ are negative in all cases
whereas the low-frequency mobile carrier contribution to $\chi^{(3)}$ for 
bulk InAs is positive.  
It is found that $\chi^{(3)}_{zzzz}(0)$ is non-zero even for the parabolic 
band approximation, which suggests that nonlinearity is mainly due to 
confinement of electron wave function within the quantum well region. When 
nonparabolicity is included, the dipole matrix elements do not change much
\cite{war2}, but the zone centre ISBT energies decrease compared to those 
for the parabolic band approximation, and continue to decrese with increase 
in the in-plane momentum. As a result, both the terms within the 
parenthesis in Eq. \ (\ref{chi30}) increase. But, the rate of increase 
of the first term which is smaller of the two terms, is more than that 
of the second. Hence $|\chi^{(3)}_{zzzz}(0)|$ decreases. 

It is instructive to compare these results with the values of 
$\chi^{(3)}_{zzzz}(0)$ calculated for infinitely deep QWs using 
Eq. \ (\ref{chi0}). We see that the value of $|\chi^{(3)}_{zzzz}(0)|$ is
much larger than that obtained by using Eq. \ (\ref{chi0}) for infinite 
barrier model. This is expected for 
two reasons. Firstly, since a real-life QW system has a finite barrier 
height, the electron wavefunction penetrates into the barrier region. As the 
electron spends part of the
time in the barrier layer, the effective well width increases, and
$|\chi^{(3)}_{zzzz}(0)|$ being proportional to the ninth power of the well
width, for finite barrier heights, its value increases by an order of 
magnitude even for the parabolic band approximation. Secondly, in the 
calculation using Eq. \ (\ref{chi0}) partially occupied subbands are not
included, i.e. only the transitions at or near the zone centre are considered.
This precludes  transitions from the first to the second 
subband when the latter is populated. However, such transitions do occur
\cite{war} for $k_{f2} < k_{\perp} < k_{f1}$ where $k_{f1}$ and $k_{f2}$ are 
the Fermi wave vectors for the first two subbands. Such transitions have been 
included in Eq. \ (\ref{chi30}),
which leads to an increase in the value of $|\chi^{(3)}_{zzzz}(0)|$.

From these results we may conclude that nonlinearity due to ISBT is influenced
by both the band nonparabolicity as in bulk semiconductors, and the quantum
confinement of the electron wavefunction. The main contribution is from the
quantum confinement which is substantially modified by nonparabolicity. 

\section{Third harmonic generation, $\chi^{(3)}_{\lowercase{zzzz}}(3\omega)$ }

The real and the imaginary parts of $\chi^{(3)}_{zzzz}$ due to THG are shown 
in Fig 1. The calculation have been done for InAs/AlSb MQW system by sum-over
-states method. There are 14 bound levels, all the levels are taken into
account in calculations. It is seen that the value $\chi^{(3)}(3\omega)$ 
converges after taking 10 bound levels. 
Because sum of the oscilator strengths for the bound states
becomes close to unity, the effect of continuum may be safely neglected.
In Fig 2. $|\chi^{(3)}_{zzzz}(3\omega)|$ is shown,
which is a measurable quantity. There are 8 peaks due to $\omega$,
2$\omega$, and 3$\omega$ resonances from the two occupied subbands. The
$\omega$ , 2$\omega$, and 3$\omega$ resonances due to transitions from the
ground subband occur at $\hbar \omega = $ 77, 95, 25, and 105 meV, 
respectively. The resonances due to transitions from the first
excited level occur at $\hbar \omega = $ 110, 117, 36, and 120 meV, 
respectively. The peaks at 25 and 36 meV are very small, as shown 
in the figure. It is seen that around 117 meV, i.e., corresponding
to the 10.6 $\mu$m region, a very high value, about  $4.4 \times 10^{-4}$ 
e.s.u. of $|\chi^{(3)}_{zzzz}(3\omega)|$ is obtained, which is the 
largest value of nonlinearity due to THG ever reported for a single 
symmetric QW system. Since this MQW system already exists, it is worthwhile
to do experiments to establish high nonlinearities of purely electronic 
origin. We note that when depolarization effects are included, one-photon
resonances would shift from the calculated positions as in one photon spectra.

\begin{figure}
\psfig{file=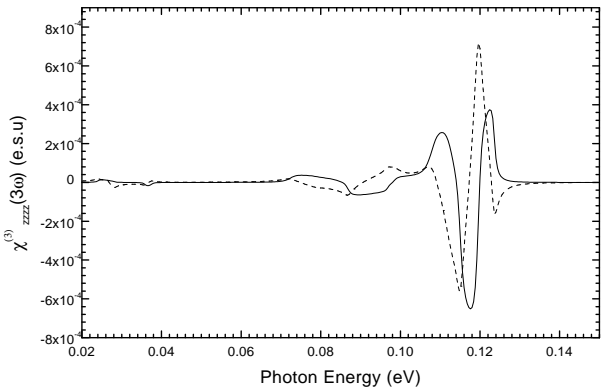,width=9cm}
\caption{ Dependence of $\chi^{(3)}(3\omega)$ on photon energy for the
InAs/AlSb MQW system. The dotted and the solid lines show the real and the
imaginary parts, respectively.}
\end{figure}

\begin{figure}
\psfig{file=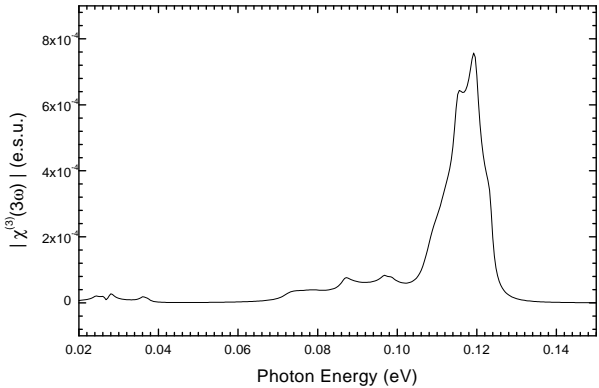,width=9cm}
\caption{ Dependence of $|\chi^{(3)}(3\omega)|$ on the photon energy for the
InAs/AlSb MQW system.}
\end{figure}

\section{Nonlinear refraction and absorption, $\chi^{(3)}_{\lowercase{zzzz}}
(-\omega)$}

The real and the imaginary parts of $\chi^{(3)}_{zzzz}(-\omega)$ are shown in
Fig. 3. The imaginary part describes TPA.
Here we obtain four major peaks due to transition from the ground and
the first excited state. 
\begin{figure}
\psfig{file=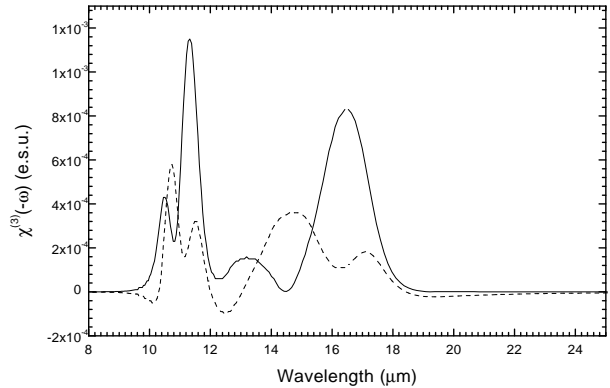,width=9cm}
\caption{ Dependence of $\chi^{(3)}(-\omega)$ on the incident wavelength
for the InAs/AlSb MQW system. The dotted and the solid lines show the real 
and the imaginary parts, respectively.}
\end{figure}
From the figure it is seen that at about
10.6 $\mu$m there is a resonance due to two photon absorption between the 
second and the fourth subbands. The magnitude of third order optical 
nonlinearity corresponding to this TPA is also very high, of about 4 
$\times $ 10$^{-4}$ e.s.u. It is comparable to the largest reported 
nonlinearity at
10.6 $\mu$m for Hg$_{0.84}$Cd$_{0.16}$Te (Ref. \onlinecite{you}). It should 
be noted that this resonant nonlinearity is of practical interest because no 
linear absorption occurs at this wavelength.
For all optical switching devices the ratio, $\Bigg| \frac{Re \left( 
\chi^{(3)} \right)}{Im \left( \chi^{(3)} \right)} \Bigg|$ representing the 
phase of the nonlinearity is important, and should have a value greater than 
2 (Ref. \onlinecite{khu2,prlpap}) for device applications. This ratio is 
plotted in Fig. 4, and we see that the condition is satisfied near the TPA
wavelength. We have earlier predicted \cite{prlpap} intrinsic optical
bistability in this sample considering electron-electron interaction through
effective medium theory.

The real and the imaginary parts of $\chi^{(3)}_{zzzz}(-\omega)$ of 
GaAs/GaAlAs MQW system are plotted in Fig. 5. The physical parameters used in
the calculations correspond to the sample used by Craig et al. \cite{cra}
are listed in Table I. 

The results for
the higher carrier concentration ($N_s = 1.1 \times 10^{11}$ cm$^{-2}$) are
given in this figure. Since only one energy level is populated here, two 
peaks, one each for one photon and two photon resonances are observed. The 
high value of $\chi^{(3)}(-\omega)$ (2.2 $\times 10^{-2}$ e.s.u.) 
at the TPA frequency is due to the large well width of 40 nm and high
carrier concentration.
Since $\Bigg| \frac{Re \left( \chi^{(3)} \right)}{Im \left( \chi^{(3)} 
\right)} \Bigg| > 2$ (see Fig. 6) near the TPA frequency, this particular QW 
system is also expected to be useful for optical switching. 

\begin{figure}
\psfig{file=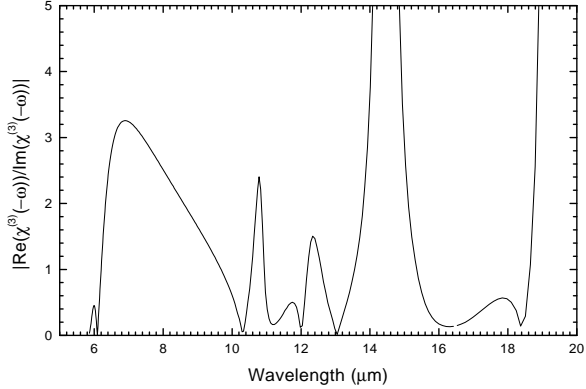,width=9cm}
\caption{ Dependence of $\Big| Re \left( \chi^{(3)}(-\omega) \right)/
Im \left( \chi^{(3)}(-\omega) \right) \Big|$ on the incident wavelength for 
the InAs/AlSb MQW system.}
\end{figure}
\begin{figure}
\psfig{file=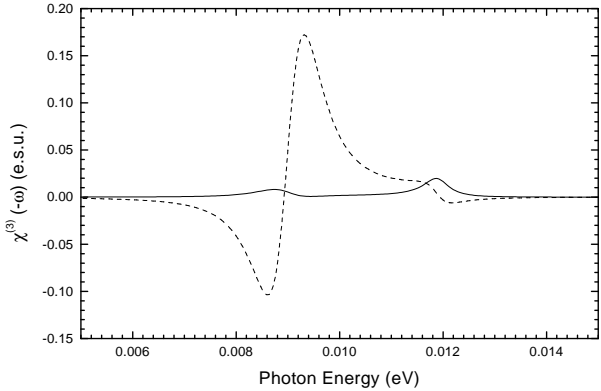,width=9cm}
\caption{ Dependence of $\chi^{(3)}(-\omega)$ on the photon energy
for the GaAs/GaAlAs MQW system. The dotted and the solid lines show the real 
and the imaginary parts, respectively.}
\end{figure}

Using Eq. \ (\ref{chi3tpa}) we have also calculated  
$|\chi^{(3)}_{zzzz}(-\omega)|$ for GaAs/GaAlAs MQW systems used by Walrod
et al. \cite{wal} and obtained a set of values of 3.65 $\times$ 10$^{-5}$ and 
3.91 $\times$ 10$^{-5}$ e.s.u. which are in good agreement with their
experimental results, 3.3 $\times$ 10$^{-5}$ and 4.25 $\times$ 10$^{-5}$ 
e.s.u., respectively.
\begin{figure}
\psfig{file=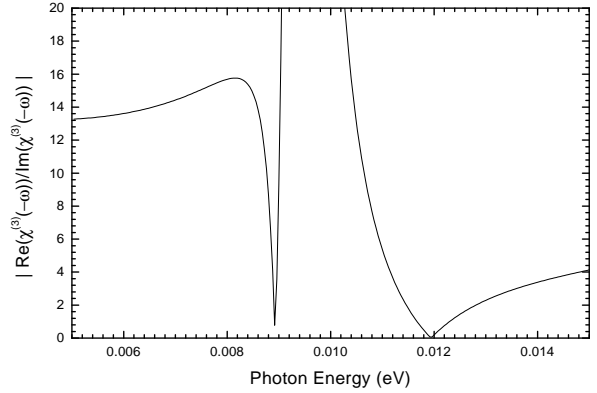,width=9cm}
\caption{ Dependence of $\Big| Re \left( \chi^{(3)}(-\omega) \right)/
Im \left( \chi^{(3)}(-\omega) \right) \Big|$ on the photon energy for the 
GaAs/GaAlAs MQW system.}
\end{figure}

\section {intensity-dependent absorption}

To calculate the intensity-dependent absorption we consider the GaAs/AlGaAs
system used by Craig et al. \cite{cra} The values of transition electric 
dipole moments and the dephasing relaxation time, $T_2$ are taken from the
experimental work \cite{cra}. We also assume $T_1 = T_2$. In the effective 
medium theory the 
intensity in, and the intensity-dependent dielectric function of the well 
region depend on each other through Eqs. \ (\ref{2level}) and (\ref{eweb}). 
These were 
solved self-consistently at each intensity to obtain $\tilde{\epsilon}$ whose
imaginary part determines the absorption by the composite medium. These 
results are shown in Fig. 7. 
\begin{figure}
\psfig{file=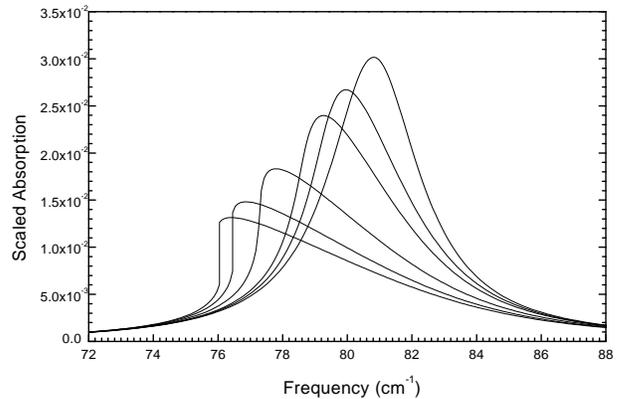,width=9cm}
\caption{ Dependence of scaled absorption on the frequency of the incident
radiation for the GaAs/GaAlAs MQW system for $T_1=T_2$. The intensities of
the radiation for the highest to the lowest peaks are $10^{-3}$, 100, 200,
500, 800, and 1000 Wcm$^{-2}$, respectively.}
\end{figure}
Here, the absorption co-efficient, multiplied 
by a scale factor, $\hbar n_0 c/q$ where $n_0$ $c$, and $q$ are the 
refractive index of bulk GaAs, velocity of light, and the electronic charge, 
respectively is plotted in the ordinate. The asymmetry in the lineshapes
of the absorption curves is known to occur due to the collective effect 
\cite{zal} and is well reproduced in our calculation. We note
that the observed position of the linear absorption peak at 81 cm$^{-1}$ is 
blue shifted from the calculated one electron peak position at 72 cm$^{-1}$,
which can be explained as the depolarization shift with the help of the many
electron theory. This shift has also been accurately reproduced in our 
calculations. Using an intensity-dependent $T_1$ it is possible to 
get much better agreement 
with the experimental intensity-dependent absorption spectrum  for the 
sample having $N_s = 1.1 \times $ 10$^{11}$ cm$^{-2}$. 
For the other samples, a change in the absorption frequency by
5 cm$^{-1}$ is observed for high incident intensities. The discrepancy is
due to the asymmetry in the QW shape caused by an applied DC electric field
which has not been included in our calculations. We have also
obtained similar results for the GaAs/GaAlAs system discussed by Zaluzny 
\cite{zal}.

\section{conclusions}
In this paper we have presented detailed calculations of the third order 
optical nonlinearity due to ISBT in semiconductor MQW systems. We first
calculate the single particle linear and nonlinear response functions. The
most important effects of electron-electron interactions i.e. the screening 
due to depolarization terms are included then through an effective medium 
theory. In calculating the single particle response all the 
essential complexities, such as the finite barrier height and the energy band 
nonparabolicity are included. Very high values of
$\chi^{(3)}_{zzzz}(-\omega)$ ($4 \times 10^{-4}$ e.s.u.) and 
$\chi^{(3)}_{zzzz}(3\omega)$ ($4.4 \times 10^{-4}$ e.s.u.) are predicted
near 10.6 $\mu$m for the InAs/AlSb MQW system. We find that in the 
low-frequency regime the nonlinearity is mainly due to quantum confinement of
electrons - the effect of nonparabolicity being a modification of this 
nonlinearity. 

The inclusion of electron-electron screening terms through an effective medium
theory gives rise to several new features. In linear spectra the dominant 
absorption peaks are attributed to intersubband plasmons. The blue and 
the red shifts observed in the linear and the nonlinear optical absorptions,
respectively due to the many electron effect can be explained in terms of 
the effective medium theory. The asymmetry of the lineshapes of the 
nonlinear absorption spectra for intensities higher than the saturation 
intensity is also obtained. 

\acknowledgements
It is a pleasure to thank Dr. Kanad Mallik for useful discussions and 
suggestions.


\begin{references}
\bibitem[*]{oldaff}
The work was done when the author was a visiting scientist at the Centre for 
Advanced Technology, Indore.
\end{references}
\end{document}